\documentclass{iaus}
\usepackage{graphicx}

\title[MOST Photometry of 55 Cancri] 
{New MOST \footnote{Based on data from the MOST satellite, a Canadian Space Agency mission operated by Microsatellite Systems Canada Inc. (MSCI; former Dynacon Inc.) and the Universities of Toronto and British Columbia, with the assistance of the University of Vienna.} Photometry of the 55 Cancri System}

\author[Diana Dragomir, Jaymie M. Matthews, Joshua N. Winn et al.]   
{Diana Dragomir$^{1,2}$, Jaymie M. Matthews$^1$, Joshua N. Winn$^3$, Jason F. Rowe$^4$
 \and MOST Science Team}

\affiliation{$^1$Department of Physics and Astronomy, University of British Columbia, Vancouver, BC V6T1Z1, Canada \\ [\affilskip] $^2$Las Cumbres Observatory Global Telescope Network, 6740 Cortona Drive, Suite 102, Santa Barbara, CA 93117, USA \\ email: {\tt ddragomir@lcogt.net} \\[\affilskip] $^3$Dept. of Physics, and Kavli Institute for Astrophysics and Space Research, Massachusetts Institute of Technology, Cambridge, MA 02139, USA \\[\affilskip] $^4$NASA Ames Research Center, Moffett Field, CA 94035}

\pubyear{2012}
\volume{xxx}  
\jname{Title of your IAU Symposium}
\editors{A.C. Editor, B.D. Editor \& C.E. Editor, eds.}
\begin{document}

\maketitle

\begin{abstract}

Since the discovery of its transiting nature, the super-Earth 55 Cnc e has become one of the most enthusiastically studied exoplanets, having been observed spectroscopically and photometrically, in the ultraviolet, optical and infrared regimes. To this rapidly growing data set, we contribute 42 days of new, nearly continuous MOST photometry of the 55 Cnc system. Our analysis of these observations together with the discovery photometry obtained in 2011 allows us to determine the planetary radius (1.990$^{+0.084}_{-0.080}$ R$_{\oplus}$) and orbital period (0.7365417$^{+0.0000025}_{-0.0000028}$ days) of 55 Cnc e with unprecedented precision. We also followed up on the out-of-transit phase variation first observed in the 2011 photometry, and set an upper limit on the depth of the planet's secondary eclipse, leading to an upper limit on its geometric albedo of 0.6.

\end{abstract}

\firstsection 
              
\section{Introduction}
\label{sec:55intro}

Currently, the most straightforward method for increasing the number of known transiting super-Earths around bright stars is to monitor the host stars of RV-discovered exoplanets with minimum masses in the super-Earth regime during their predicted times of inferior conjunction relative to the Earth. The planet 55 Cnc e is the first super-Earth whose transiting nature was discovered in this way. The host star of this system is a very bright ($V=5.95$) G8 dwarf, and 55 Cnc e orbits it in only 0.74 days. The transit detection was achieved independently with the MOST space telescope (\cite[Winn et al. 2011]{Winn3}) and with the NASA Spitzer infrared space telescope detection (\cite[Demory et al. 2011]{Dem11}).

Beyond the transit detection, both instruments have revealed further properties of this system and its innermost, transiting planet. The MOST light curve hinted at out-of-transit variations in phase with 55 Cnc e's orbit. The modulation, whose amplitude is too large to be due to reflected light from 55 Cnc e, may instead indicate the existence of tidal or magnetic interactions between the planet and its host star. Spitzer was employed to observe four occurrences of the planet's secondary eclipse using its 4.5 $\mu$m channel, from which a brightness temperature of 2360 $\pm$ 300 K was determined (\cite[Demory et al. 2012]{Demory2}). The authors propose that the measured brightness temperature may suggest inefficient heat redistribution combined with a low albedo, but that it is also possible that the 4.5 $\mu$m observations probe a deeper layer in the planet's atmosphere, thus potentially indicating the presence of a thermal inversion layer. Recently, \cite[Crossfield (2012)]{Crossfield} obtained NASA InfraRed Telescope Facility (IRTF) spectroscopic observations of 55 Cnc A and derived an improved flux density for the star, leading to a re-estimate of the Spitzer data-based brightness temperature of 55 Cnc e. From these measurements, he obtains a slightly lower temperature of 1950$^{+260}_{-190}$ K for the planet, which expands the range of possible albedo values consistent with this temperature. 

In this proceedings, we describe the acquisition and reduction of new MOST observations aimed at searching for the secondary eclipse of 55 Cnc e, verifying whether the previously observed phase variations persist, and improving the physical and orbital parameters of this transiting super-Earth in Section~\ref{sec:55obs}. 

\section{Observations and Data Reduction}
\label{sec:55obs}

55 Cnc was observed nearly continuously using the MOST space telescope for about 42 days from 2012 January 14 to February 25. The observations were acquired in Direct Imaging mode with an exposure time of 0.5 s per individual frame. The images were downloaded from the satellite in stacks of 80 frames, resulting in a total integration time of 40 s per data point. 

A raw light curve was extracted from the images using aperture photometry. The magnitudes were fitted to a linear function of sky background and x-position polynomials. 

\begin{figure}[!h]
\begin{center}
\includegraphics[scale=0.3]{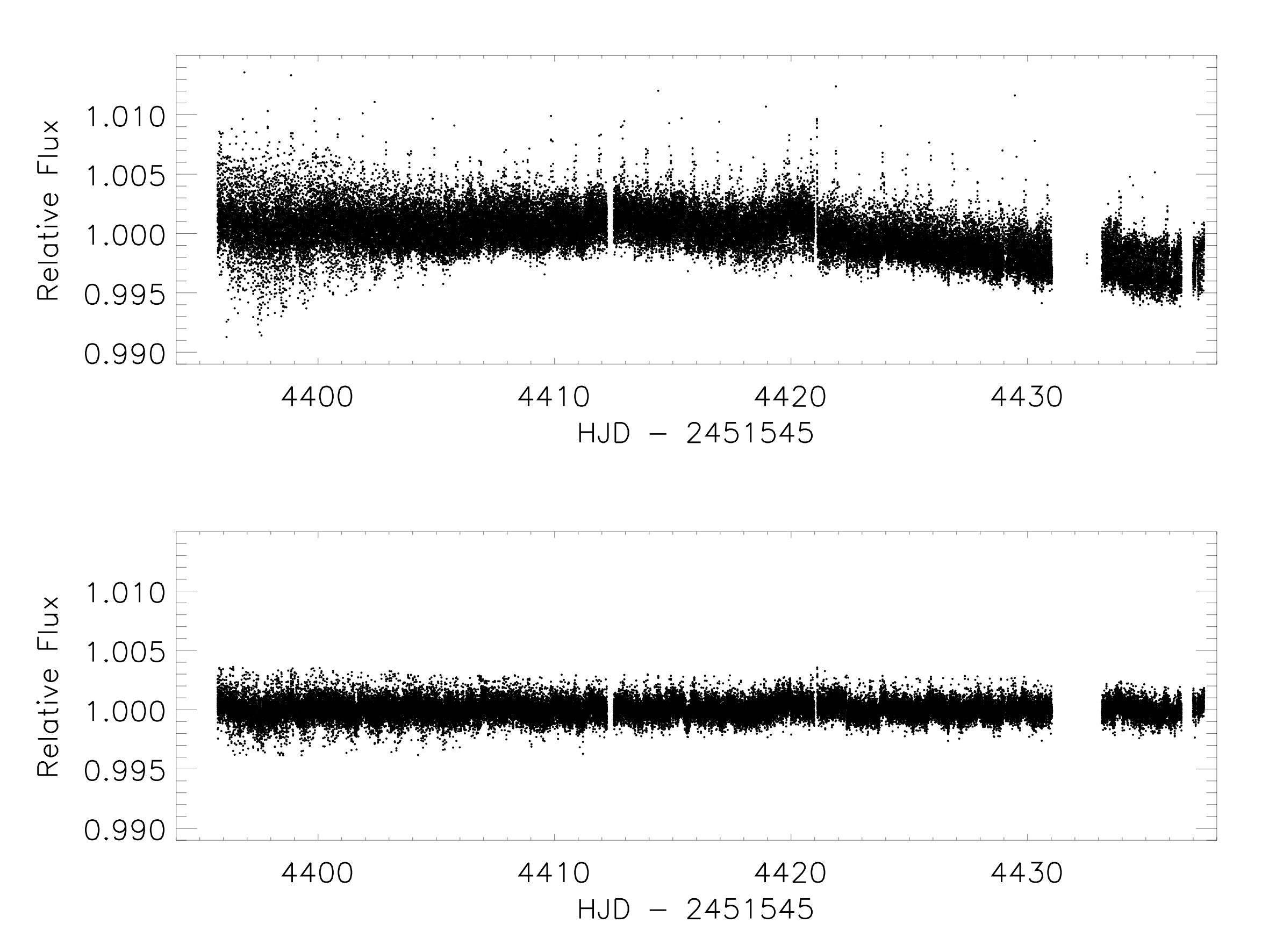}
\caption[55 Cnc MOST 2012 photometry]{\label{fig:55all}MOST 2012 light curve after de-correlation of magnitudes from sky background and x and y pixel position, and the running average correction for straylight variations at the MOST orbital period.}
\end{center}
\end{figure}

The remaining routine correction is the filtering of straylight variations modulated with the satellite's orbit of 101.4 minutes. This was accomplished by dividing the light curve into 21 sections approximately 2 days long each, folding each section at the satellite's orbital period and computing a running average of this phased data. We then removed the resulting waveform from that section of light curve. The final light curve contains 62 337 points and is shown in the bottom panel of Figure~\ref{fig:55all}.

\section{Analysis of the Light Curve}
\label{sec:55ana}

We used a Markov Chain Monte Carlo (MCMC) algorithm to analyze the MOST photometry with the goals of improving the values of the transit and orbital parameters of 55 Cnc e and searching for the out-of-transit variation originally observed in 2011 (\cite[Winn et al. 2011]{Winn3}) as well as the secondary eclipse. For model comparison, we use the odds ratio. The Bayesian Information Criterion (BIC) is used as an estimate of the marginal likelihood for each model (\cite[Carlin \& Louis 2009]{Carlin}). 

We fitted the following eight parameters: the orbital period ($P$), the mid-transit time ($T0$), the ratio of the planet to star radii ($R_{P}/R_{\star}$), the scaled semi-major axis ($a/R_{\star}$), the orbital inclination ($i$), the depth of the secondary eclipse ($\delta$), the amplitude of the orbital phase modulation assuming a sinusoidal function with its minimum coinciding with the mid-transit time ($\alpha$) and a normalisation factor corresponding to the flux just outside of transit ($F_{norm}$). Including quadratic limb-darkening coefficients in our fit did not improve the odds ratio, suggesting the limb-darkening is not well constrained by our photometry. These two parameters were thus held fixed at $u_1=0.648$ and $u_2=0.117$, the MOST bandpass values for the grid point of stellar parameters ($T_{eff}=5200$ K, $\log g =4.5$, $[Fe/H] = 0.2$; A. Prsa, private communication) nearest to those of 55 Cnc. Gaussian priors were used for $P$ and $a/R_{\star}$, and uniform priors for the remaining floating parameters.

We carried out a MCMC analysis of the combined 2011 and 2012 light curves. The results of this analysis are listed in column 1 of Table~\ref{tab:55cncprops2}. While we do not detect a significant non-zero secondary eclipse, we are able to use the best-fitted value of $\delta$ and its uncertainties to place an upper limit on its depth. Because fitting for the secondary eclipse does not improve the model likelihood, we also ran the MCMC analysis omitting this parameter in order to obtain improved uncertainties on the remaining parameters. The results can be seen in column 2 of Table~\ref{tab:55cncprops2}.

\begin{figure}[!h]
\begin{center}
\includegraphics[scale=0.32]{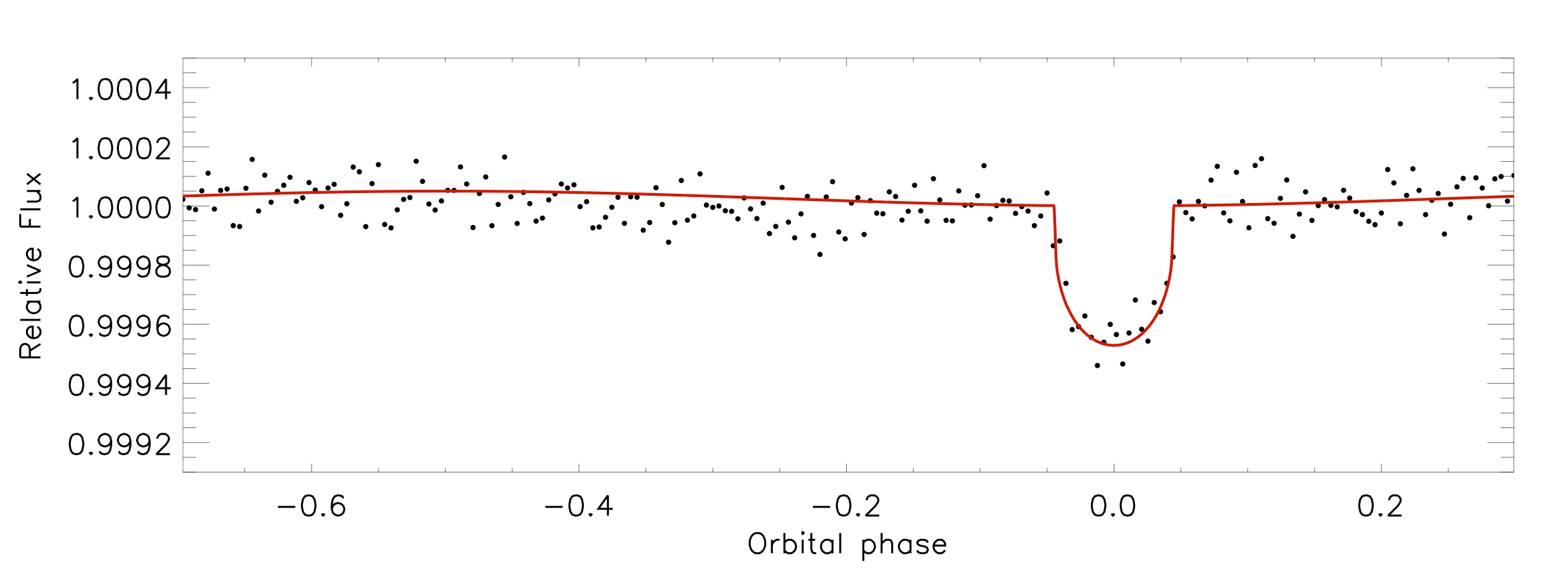}
\caption[Phase curves for 55 Cnc e]{\label{fig:55Cnce} 2011 and 2012 photometry phased at the orbital period of 55 Cnc e and averaged in 5-min phase bins. The red line is the best-fitting transit model without a secondary eclipse parameter, based on the values in column 2 of Table~\ref{tab:55cncprops2}. See Section~\ref{sec:55ana} for details.}
\end{center}
\end{figure}

For all parameters but $\delta$ the column of interest is therefore column 2. The combined photometry phased at the best-fitted period of 55 Cnc e is shown in the Figure~\ref{fig:55Cnce}, with the best fit model based on the parameters listed in column 2 superimposed. The orbital period we obtain is more precise than any published values for 55 Cnc e, and it is in good agreement with those values (\cite[Dawson \& Fabrycky 2010]{Dawson}). Our transit parameter values agree with those of \cite[Winn et al. (2011)]{Winn3} within the uncertainties, though only marginally for $i$ and $a/R_{\star}$. In particular, our non edge-on best-fitted orbital inclination value has implications for the transit probabilities of the other planets in the system (\cite[Fischer et al. 2008]{Fischer1}), if their orbits are co-planar with that of 55 Cnc e to within a few degrees. 

\begin{table}
\begin{center}
\caption{Orbital and transit parameters for 55 Cnc \lowercase{e} (2)}
\label{tab:55cncprops2}
{\scriptsize
 \begin{tabular}{|l|c|c|c|c|}\hline 
\textbf{Parameter} &  \textbf{2011 + 2012 photometry} & \textbf{2011+2012 photometry}  \\
\textbf{} & \textbf{(with secondary eclipse)} & \textbf{(without secondary eclipse)}  \\
\hline
\textbf{Measured parameters} & & \\
  Period - $P$ (days)                        & $0.7365418^{+0.0000027}_{-0.0000026}$ & ${\bf 0.7365417^{+0.0000025}_{-0.0000028}}$ \\
  Mid-transit time - $T_0\,^{a}$ (BJD)  & $2455962.0693^{+0.0017}_{-0.0018}$ & ${\bf 2455962.0697^{+0.0017}_{-0.0018}}$ \\
  Planet-to-star radii ratio - $R_p/R_{\star}$ & $0.01929^{+0.00074}_{-0.00077}$ & ${\bf 0.01936^{+0.00079}_{-0.00075}}$ \\
  Transit depth - $(R_p/R_{\star})^2$ (ppm) & $372^{+29}_{-30}$ & ${\bf 374^{+31}_{-29}}$ \\
  Scaled semi-major axis - $a/R_{\star}$  & $3.524^{+0.041}_{-0.042}$ & ${\bf 3.523^{+0.042}_{-0.040}}$  \\
  Orbital inclination - $i$ (deg)  & $85.5^{+2.8}_{-2.0}$ & ${\bf 85.4^{+2.8}_{-2.1}}$  \\
  Amplitude of phase variation - $\alpha$ (ppm) & $-$ & ${\bf 34^{+12}_{-11}}$ \\
  Eclipse depth - $\delta$ (ppm)  & $-1^{+18}_{-22}$ & ${\bf -}$ \\
  & & \\ \hline
  \textbf{Derived parameters} & & \\
  Transit duration - $t_d$ (days) & $0.0662^{+0.0034}_{-0.0026}$ & $0.0660^{+0.0035}_{-0.0028}$ \\ 
  Semi-major axis - $a$ (AU)  & $0.01545^{+0.00024}_{-0.00025}$ & ${\bf 0.01545^{+0.00025}_{-0.00024}}$ \\
  Planetary radius - $R_p$ ($R_{\oplus}$) & $1.983^{+0.079}_{-0.082}$ & ${\bf 1.990^{+0.084}_{-0.080}}$  \\ \hline
  \end{tabular}
}
\end{center}
\vspace{1mm}
\end{table}


\section{Discussion}
\subsection{The Radius and Composition of 55 Cnc e}
\label{sec:55rad}

Our analysis of all the available MOST photometry of the system results in a planetary radius of $1.990^{+0.084}_{-0.080} R_{\oplus}$. Our new estimate, and the MOST  2011 (\cite[Winn et al. 2011]{Winn3}) and Spitzer (\cite[Gillon et al. 2011]{Gillon2}) values agree with each other within their uncertainties. We have determined the radius of 55 Cnc e (as measured in a broad optical passband) with a precision of 4\%. We used the planetary mass derived by \cite[Demory et al. (2011)]{Dem11} ($M_p=7.81 \pm 0.56 M_{\oplus}$) to obtain the mean density, and obtained a value of 5.489$^{+0.77}_{-0.80}$ g cm$^{-3}$, one of the best constrained for a super-Earth. The 1$\sigma$ upper limit on its density just reaches a composition in the coreless silicate regime (\cite[Elkins-Tanton \& Seager]{Elkins}), but the span of the uncertainty lies mainly within the volatile envelope planets area of theoretical mass-radius models. 

\cite[Gillon et al. (2011)]{Gillon2} examined two plausible compositions which match their mass and radius measurements for 55 Cnc e: an envelope consisting of 0.1\% H/He by mass and one consisting of 20\% H$_2$O by mass. They argue, based on the calculations of \cite[Valencia et al. (2010)]{Valencia4}, that the former would evaporate within a few million years. A water vapour envelope would also be depleted over time, but only very slowly; the process is expected to take several billion years. Given the estimated age of the star ($10.2 \pm 2.5$ Gyr; \cite[von Braun et al. 2011]{Braun3}), a water vapour atmosphere is by far the more likely scenario. Our value for the radius of 55 Cnc e is consistent with this model. 

\subsection{Brightness Modulation at the Planetary Orbital Period}
\label{sec:55oot}

\cite[Winn et al. (2011)]{Winn3} suggested the phase modulation discovered in the 2011 MOST photometry, too large to be explained by scattered light from the planet, could be attributed to star-exoplanet magnetic interaction. Because the orbital motion of the planet and the stellar rotation are likely not synchronised (the best estimate of the stellar rotation period is $\sim$43 days; \cite[Fischer et al. 2008]{Fischer1}) , we do not expect interaction manifesting itself as a region of star spots and chromospheric activity fixed in the co-rotating frame of the star at the sub-planetary point, but rather a travelling wave moving through the photosphere and chromosphere of the star. Such an effect can naturally be expected to have a changing amplitude, and in fact that is what \cite[Gillon et al. (2011)]{Gillon2} found in their re-analysis of the MOST 2011 photometry. 

\begin{figure}[!h]
\begin{center}
\includegraphics[scale=0.32]{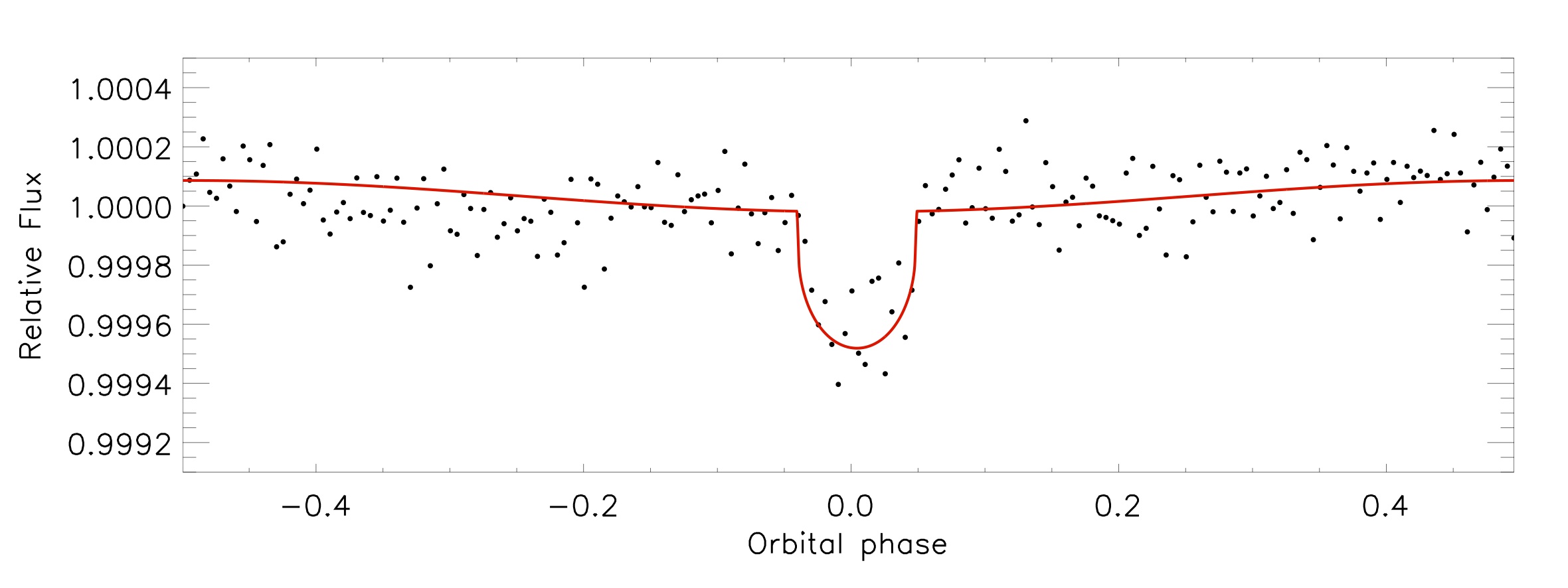}
\caption[Phase curve for 55 Cnc e based on part of the 2012 photometry]{\label{fig:55part} Approximately 2/3 of 2012 light curve phased at the orbital period of 55 Cnc e and averaged in 5-min phase bins. The red line is the best-fit transit model without a secondary eclipse parameter, and shows a more pronounced phase variation than when the complete 2012 light curve is used (see Section~\ref{sec:55oot} for details).}
\end{center}
\end{figure}

When folding the entire 2012 MOST light curve, we do not find a phase variation at the period of 55 Cnc e as strong as that present in the 2011 light curve. This motivated us to inspect the 2012 data set more closely and, while always smaller than in the 2011 data, the amplitude of the phase variation does appear to change over the course of the observations. Interestingly, if we fold in phase the photometry only between HJD 2455945 (near the end of the portion of the light curve more affected by stray light modulation) and HJD 2455969, we find the amplitude of the phase variation ($42 \pm 16$ ppm) is about twice what we find by folding the entire data set. This suggests that the amplitude of the variation varies on a timescale of weeks and perhaps even days, which is consistent with the finding of \cite[Gillon et al. (2011)]{Gillon2}. This section of the light curve (phased and binned as before) is shown in Figure~\ref{fig:55part}, with the best-fitting model superimposed.

\subsection{Eclipse and Albedo Limits}
\label{sec:55alb}

We have not detected a secondary eclipse for 55 Cnc e. Our best estimate for the depth of the eclipse is $\delta=-1^{+18}_{-22}$ ppm, and we can use this result to set an upper limit of $\delta \leq$ 17 ppm. 

We can express the ratio of the planet-to-star fluxes ($F_p/F_{\star}$) as a function of the geometric albedo $A_g$ (the fraction of light reflected or scattered by the planet at optical wavelengths), planetary radius and semi-major axis:

\begin{eqnarray}
\frac{F_p}{F_{\star}} = A_g (\frac{R_p}{a})^2
\end{eqnarray}

\noindent A geometric albedo of 1 (100\% reflective) means the eclipse could be no deeper than 30 ppm. Our upper limit on $\delta$ leads to an upper limit (1$\sigma$) of 0.57 on the geometric albedo. 

Since we do not have information on the planet's brightness variation throughout its orbit, we assume it has the reflective properties of a Lambertian surface. For a Lambert sphere, the Bond albedo is $A_B=1.5*A_g$. This relation leads to an estimate of 0.85 for the 1$\sigma$ upper limit on the Bond albedo of 55 Cnc e, in agreement with the results of \cite[Demory et al. (2012)]{Demory2}.

If the surface of 55 Cnc e is molten and assuming the absence of an atmosphere (still possible if the planet is coreless and made of pure rock) or the presence of a transmissive atmosphere, the planet's geometric albedo is predicted to be 0.6 (\cite[Kane et al. 2011]{Kane2}). With our measurement of the geometric albedo, we can exclude this model with 1$\sigma$ confidence.


\section{Summary}
\label{sec:55conc}

We have acquired 42 days of new MOST photometry of 55 Cnc. After a careful reduction, we have used these data for several different types of studies. The new photometry has allowed the radius of 55 Cnc e ($R_p = 1.990 R_{\oplus}$) to be determined to within 4\%. Our result agrees with the hypothesis of \cite[Gillon et al. 2011]{Gillon2} that the planet likely has an envelope of high mean molecular weight volatiles such as water vapour and/or silicate clouds. We find that the phase variation observed in the 2011 MOST photometry persists but with an amplitude at most half of that measured in the discovery photometry. The data also allows us to place a limit on the depth of the secondary eclipse, and thus on the geometric albedo of 55 Cnc e of 0.57. This leads to an upper limit of 0.85 on the Bond albedo of the planet. With this value, we can exclude a highly reflective molten surface with 1$\sigma$ confidence. MOST photometry to be acquired in winter 2013 will bring this value into the realm of reflective silicate clouds, allowing us to place more constraining limits on the existence of such clouds in the atmosphere of 55 Cnc e.


\end{document}